\documentclass[apj]{emulateapj}
\shorttitle{ionization vs fragmentation}
\shortauthors{Zhen et al.}

\begin{document}

\title{VUV photo-processing of PAH cations: quantitative study on the ionization versus fragmentation processes}

\author{Junfeng Zhen$^{1,2}$, Sarah Rodriguez Castillo$^{1,2,3}$, Christine Joblin$^{1,2}$, \\Giacomo Mulas$^{1,2,4}$, Hassan Sabbah$^{1,2}$, Alexandre Giuliani$^{5,6}$, \\Laurent Nahon$^{5}$, Serge Martin$^{7}$, Jean-Philippe Champeaux$^{8}$, Paul M. Mayer$^{9}$}

\affil{$^{1}$Universit\'e de Toulouse, UPS-OMP, Institut de Recherche en Astrophysique et Plan\'etologie, Toulouse, France}
\affil{$^{2}$CNRS, IRAP, 9 Av. Colonel Roche, BP 44346, 31028, Toulouse Cedex 4, France}
\affil{$^{3}$Laboratoire de Chimie et Physique Quantiques LCPQ/IRSAMC, Universit\'e de Toulouse (UPS) and CNRS, \\
118 Route de Narbonne, 31062 Toulouse, France}
\affil{$^{4}$Istituto Nazionale di Astrofisica - Osservatorio Astronomico di Cagliari, via della Scienza 5, 09047 Selargius (CA), Italy}
\affil{$^{5}$Synchrotron SOLEIL, LÕOrme des Merisiers, 91192 Gif sur Yvette Cedex, France}
\affil{$^{6}$INRA, UAR1008 Caract\'erisation et Elaboration des Produits Issus de l'Agriculture, 44316 Nantes, France}
\affil{$^{7}$Institut Lumi\`ere Mati\`ere, Universit\'e Lyon 1-CNRS, Universit\'e de Lyon, 69622 Villeurbanne cedex, France}
\affil{$^{8}$Laboratoire Collisions Agr\'egats R\'eactivit\'e, Universit\'e de Toulouse, UPS-IRSAMC, CNRS,\\
118 Route de Narbonne, Bat 3R1B4, 31062 Toulouse Cedex 9, France}
\affil{$^{9}$Department of Chemistry and Biomolecular Sciences, University of Ottawa, Ottawa K1N 6N5, Canada}

\email{Correspondance author: Christine Joblin; email address: christine.joblin@irap.omp.eu }

\begin{abstract}
Interstellar polycyclic aromatic hydrocarbons (PAHs) are strongly affected by the absorption of vacuum ultraviolet (VUV) photons in the interstellar medium (ISM), yet the branching ratio between ionization and fragmentation is poorly studied.  This is crucial for the stability and charge state of PAHs in the ISM in different environments, affecting in turn the chemistry, the energy balance, and the contribution of PAHs to the extinction and emission curves.
We studied the interaction of PAH cations with VUV photons in the $7-20$~eV range from the synchrotron SOLEIL beamline, DESIRS.
We recorded by action spectroscopy the relative intensities of photo-fragmentation and photo-ionization for a set of eight PAH cations ranging in size from 14 to 24 carbon atoms, with different structures. At photon energies below $\sim$13.6~eV fragmentation dominates for the smaller species, while for larger species ionization is immediately competitive after the second ionization potential (IP). At higher photon energies, all species behave similarly, the ionization yield gradually increases, leveling off between 0.8 and 0.9 at $\sim$18~eV. Among isomers, PAH structure appears to mainly affect the fragmentation cross section, but not the ionization cross section.
We also measured the second IP for all species and the third IP for two of them, all are in good agreement with theoretical ones confirming that PAH cations can be further ionized in the diffuse ISM. 
Determining actual PAH dication abundances in the ISM will require detailed modeling. Our measured photo-ionization yields for several PAH cations provide a necessary ingredient for such models.

\end{abstract}


\section{Introduction}
\label{sec:intro}
The mid-infrared spectra of many astronomical objects are dominated by the 3.3, 6.2, 7.7, 8.6 and 11.2\,$\mu$m emission bands, which are generally attributed to the infrared (IR) fluorescence of polycyclic aromatic hydrocarbons (PAHs) that are pumped by the absorption of ultraviolet (UV) photons. These molecules contain some 10\% of the elemental carbon in space and play an important role in the ionization and energy balance of the ISM of galaxies  \citep{tie05}. Understanding the photo-chemical processes that drive the origin and evolution of these complex species and their relationship to the organic inventory of space has become a focus area in the field of astrochemistry and molecular astrophysics, as can be seen from the proceedings of dedicated conferences \citep[see e.~g.][]{PAHsUniverse}.

The astronomical PAH population is expected to evolve as a function of the local physical conditions. UV photons can lead to dissociation and ionization that are expected to compete with PAH recombination with electrons and their reactions with gas-phase species, in particular H and H$_2$. Due to this, PAHs could exist in different charge and (de)hydrogenation states depending on the environment \citep{lep01,mon13}. The presence of dications has also been considered by several authors \citep{lea86,wit06,mal07a} and predicted in some models \citep{bak01a}. As the IR spectral characteristics of PAHs are sensitive to the charge and (de)hydrogenation states, the resulting IR emission features will reflect the local physical conditions \citep[see e.~g.][]{bak01b}. Spatial variations of the IR spectrum have been observed in photodissociation regions and interpreted as due to changes in the charge from more neutral to more cationic species \citep{berne07, pil12}. Charge and (de)hydrogenation also affect the photo-absorption spectrum of PAHs \citep{cecchipestellini2008,malloci2008}, and have therefore to be considered in interstellar extinction models as well \citep[e.~g.][]{mulas2013}.

Experimental studies have so far mostly focused either on the ionization or the fragmentation aspects \citep{jochims94,jochims96,eke98,job03,west2014,zhen2014}. In fact, these two processes are in competition and both aspects have to be considered simultaneously. The recent experimental study by \citet{zhen2015} reveals that ionization dominates for the large PAH cation hexa-peri-hexabenzocoronene (HBC), C$_{42}$H$_{18}^+$, whereas it competes with fragmentation for the smaller PAH cation coronene, C$_{24}$H$_{12}^+$.
Hence, variations in the ionization characteristics of PAHs may well be as important as kinetic fragmentation parameters in chemical modeling of the interstellar PAH population, recalling the importance of including both photo-fragmentation and photo-ionization in models \citep{lep01,mon13}. 

In the present work, we perform a systematic and quantitative study on the photo-chemical behavior of PAH cations under vacuum ultraviolet (VUV) irradiation in the $7-20$ eV range. The experiments were carried out using the SOLEIL synchrotron radiation facility on eight PAHs with sizes ranging from 14 to 24 carbon atoms. The processes of ionization and fragmentation are assessed and quantified. The experimental methods are described in section~\ref{sec:exp}. The results are summarized and discussed in section~\ref{sec:results}, devoting special attention to the photo-ionization yield of PAH cations. Implication for astronomical models is presented in section~\ref{sec:discussion}.

\section{Experimental Methods}
\label{sec:exp}

\begin{figure*}[t]
  \centering
  \includegraphics[width=15 cm]{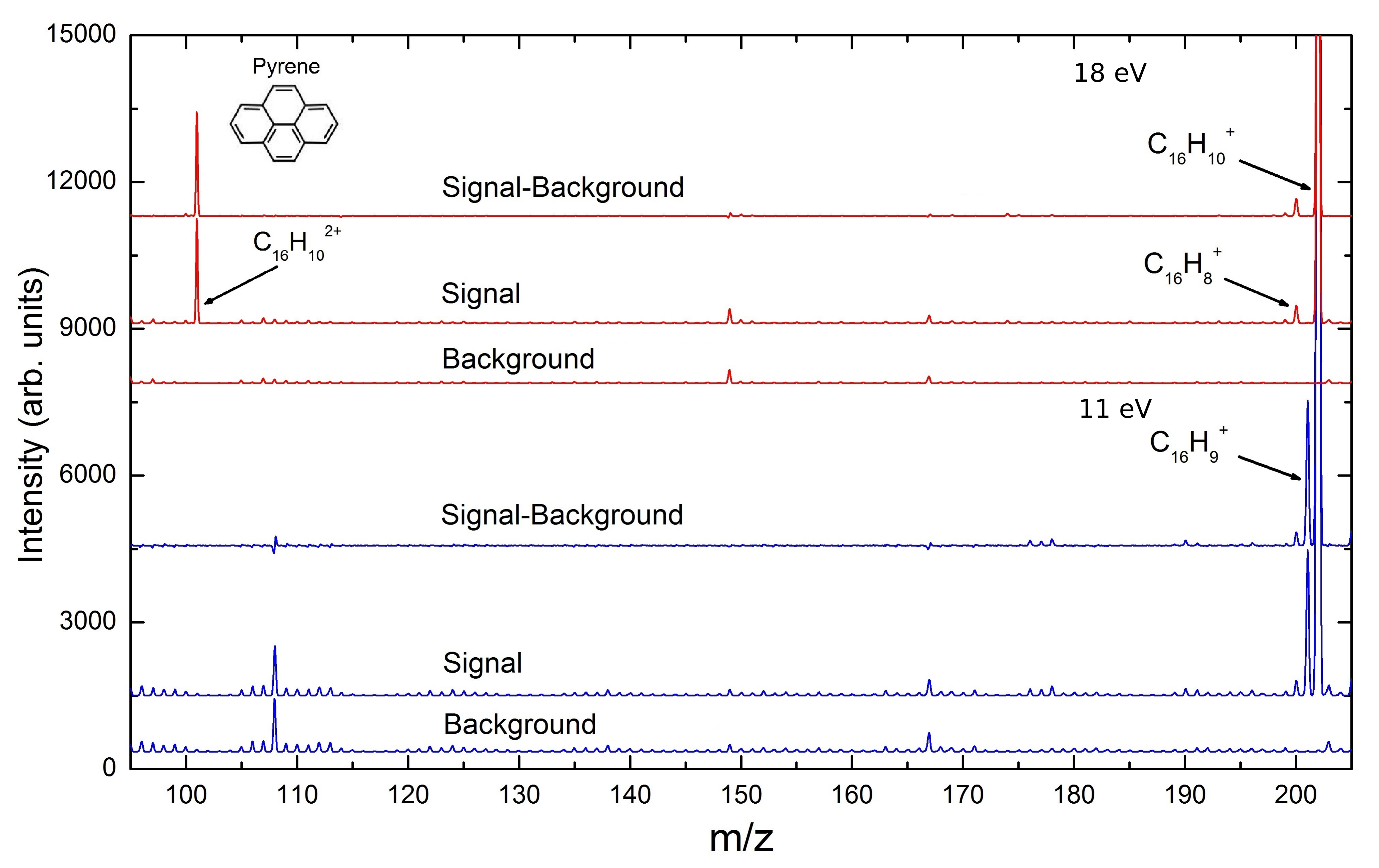}
  \caption{Example of recorded mass spectra in the case of the pyrene cation C$_{16}$H$_{10}^+$. Signal was obtained after 1.0~s of VUV irradiation at 11.0 eV (blue) and 0.2~s at 18.0~eV (red) of the parent cation. It shows the main fragmentation channels -H and -2H/-H$_2$ of the mono-cation (Fragments in Fig. ~\ref{fig_res1}~B) as well as the ionization channel leading to the formation of  C$_{16}$H$_{10}^{2+}$  and eventually to its fragments (Dication in Fig. ~\ref{fig_res1}~B). The background spectrum was recorded by switching off the ionization krypton discharge lamp.}
  \label{fig_raw_pyrene}
\end{figure*}

We have studied the photo-processing of PAH cations under VUV irradiation using action spectroscopy with the LTQ linear ion trap setup \citep{mil12} that is available on the VUV DESIRS beamline at the SOLEIL synchrotron facility in Saint-Aubin (France). The DESIRS beamline is well suited for this kind of studies because of its high photon flux, its broad and easy tunability in the 5-40 eV range, as well as its highly focused beam \citep{nah12}. We used the 6.65~m normal incidence monochromator at the first order, leading to a typical photon flux in the 10$^{12}$ $-$ 10$^{13} $ photons.s$^{-1}$ range, within a photon bandwidth of typically 12 meV at 10 eV photon energy obtained at the 200\,$\mu$m exit slit. Another crucial feature of the beamline, in this context, lies in its spectral purity; i.e., the absence of any high harmonics of the undulator, which are very efficiently filtered-out by a gas filter up to  14\,eV (krypton) and 16\,eV (argon). Above 16\,eV no filtering is needed because of the SiC coating of the normal incidence grating whose reflection curve exhibits a very sharp cut-off above 22\,eV. Absolute incident photon fluxes are measured using calibrated IRD AXUV 100 photodiodes. The VUV beam is sent directly into the linear ion trap, with a spot size of $\simeq $ 0.3 mm$^{2}$ in the center position. Irradiation time is controlled by a retractable beam shutter in the vacuum chamber.

In the LTQ linear ion trap setup, PAH cations are produced by the atmospheric pressure photo-ionization (APPI) source, { which includes nebulization of their solution in toluene and ionization} with a krypton discharge lamp producing 10.6~eV photons \citep{mor04}. This production technique is soft, no fragmentation products are observed. The species of interest (the $^{12}$C monoisotopic singly-charged parent PAH cations) are mass selected by ejecting from the trap other species (other isotopic species and fragments if they are present). Helium gas is introduced continuously directly into the ion trap. The collisions with helium atoms, at a pressure of 10 $^{-3}$ mbar, cool down the ions and improve the trapping efficiency. { We estimated that trapped PAHs experience of the order of 1000 collisions with He atoms in 100\,ms, which should be sufficient for thermalization at room temperature (293\,K). After this delay,} the ion cloud is then irradiated with the synchrotron beam, following which the photo-product pattern is mapped out. The automatic energy scanning and mass spectra recording are achieved by using a separate personal computer and a home-made program, which synchronizes the operation of both the beamline and the LTQ linear ion trap setup. 

Two sets of data were obtained for each PAH cation: one for the low incident photon energies (typically $7.0-15.6$ eV with steps of 0.3 eV) and one for the high photon energies ($14.5-20.0$ eV with steps of 0.5 eV).  For each species, an additional small scan with finer steps of 0.05 or 0.1~eV was performed around the onset of ionization, to obtain more accurate adiabatic appearance energies. Irradiation time was optimized to get the largest signal from photo-products without contamination by multiple photon absorption. An average value of 10\% for the photo-products was aimed at. This led to irradiation times of 1.0 and 0.2\,s respectively for the low and high energy ranges. Mass-spectra were then averaged around 150 (400) times for low (high) energies. In addition, blank spectra were recorded for each PAH measurement in order to subtract contaminant peaks that are also present in the background signal. The data reduction consisted then in subtracting the background and { recording the intensity for peaks in the mass spectrum above 3 sigma level from the residual (cf. Fig.~\ref{fig_raw_pyrene} as an example). The relatively high pressure of the buffer gas caused the presence of some contaminants, especially water, which may easily form complexes with PAHs and their photo-products. These complexes can be, in turn, photo-processed, leading to additional photo-products. Some care was therefore exercised in identifying out the corresponding peaks, to count them in the appropriate channels during the data analysis.

\begin{figure*}
  \centering
  \includegraphics[width=\textwidth]{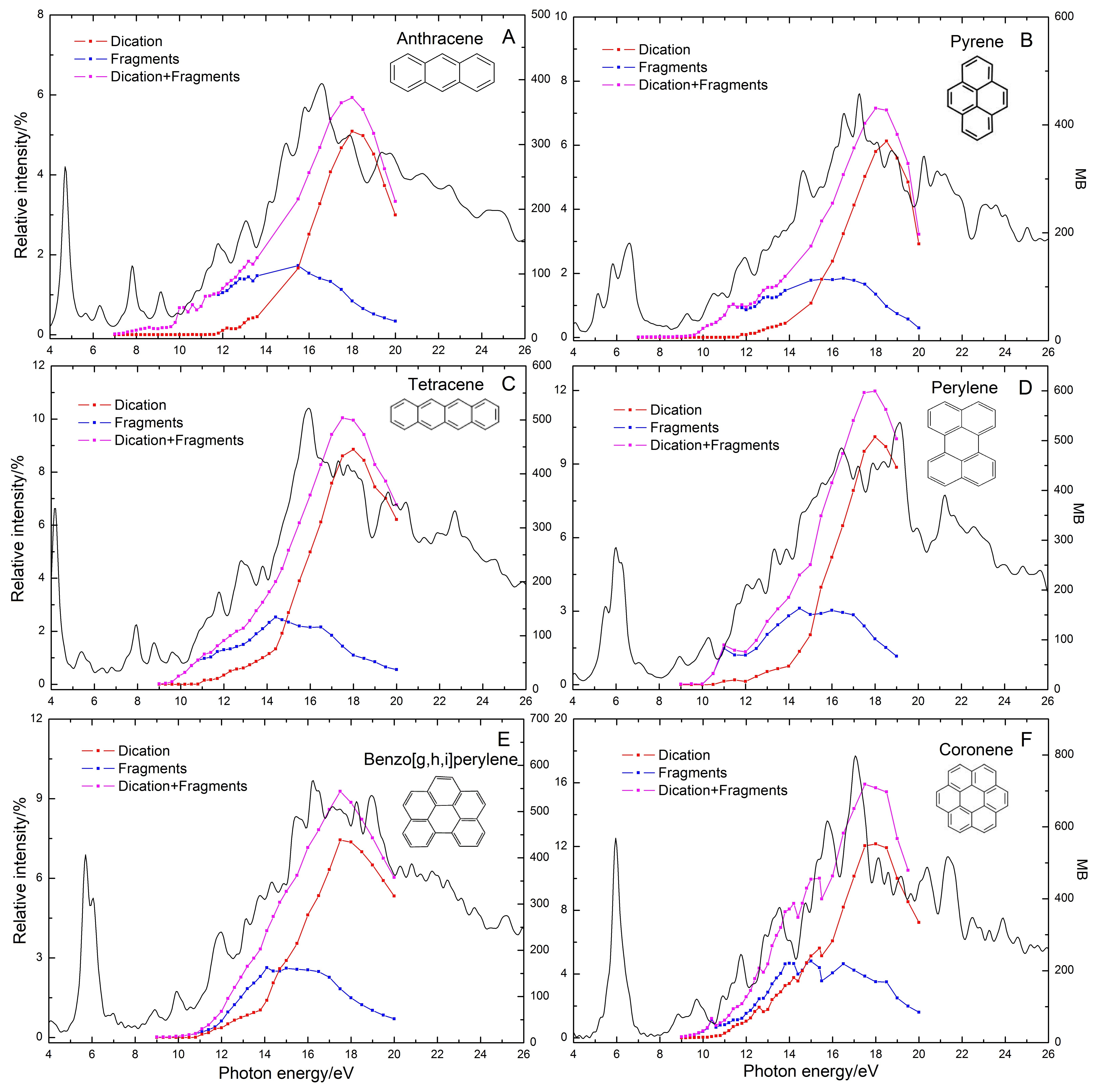}
  \caption{Values of the relative intensity of the photo-products (dications and singly-ionized fragments) of different PAHs submitted to VUV photons as a function of photon energy. The solid line is the computed photo-absorption cross-section in Mb \citep{mal04,mal07b}. Panel A: anthracene; Panel B: pyrene; Panel C: tetracene; Panel D: perylene; Panel E: benzo[g,h,i]perylene; Panel F: coronene. Irradiation time is 1.0 s for the low photon energy part (energies lower than ~15 eV) and 0.2 s for the high energy. The low energy part was scaled to the high energy part by using the two overlapping points.}
  \label{fig_res1}
\end{figure*}

\begin{figure*}
  \centering
  \includegraphics[width=\textwidth]{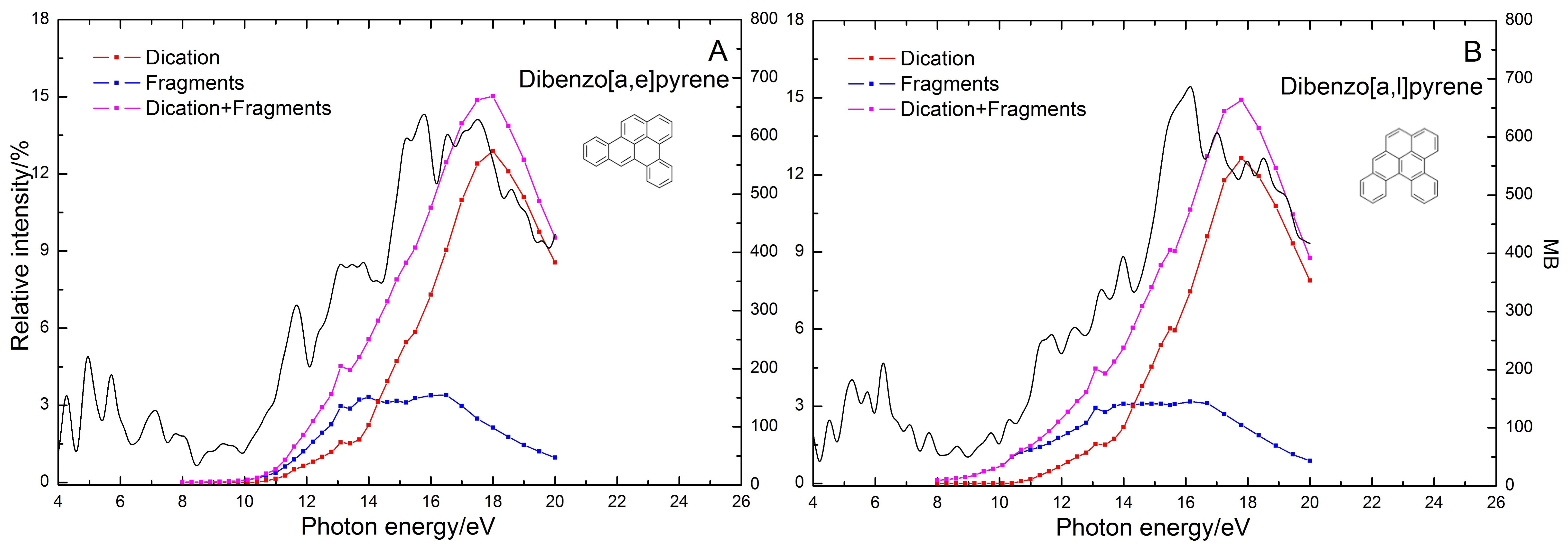}
  \caption{ Values of the relative intensity of the photo-products (dications and singly-ionized fragments) of two isomers of  dibenzopyrene submitted to VUV photons as a function of photon energy. The solid line is the computed photo-absorption cross-section in Mb (this work following \citet{mal04}). Panel A: dibenzo[a,e]pyrene; Panel B: dibenzo[a,l]pyrene.
  }
  \label{fig_res2}
\end{figure*}

\section{Results and Discussion}
\label{sec:results}
We have studied the photo-chemical processing of eight PAH mono-cations: anthracene (C$_{14}$H$_{10}^{\phantom{22}+}$), pyrene (C$_{16}$H$_{10}^{\phantom{22}+}$), tetracene (C$_{18}$H$_{12}^{\phantom{22}+}$), perylene (C$_{20}$H$_{12}^{\phantom{22}+}$), benzo[g,h,i]perylene (C$_{22}$H$_{12}^{\phantom{22}+}$), coronene (C$_{24}$H$_{12}^{\phantom{22}+}$) dibenzo[a,l]pyrene and dibenzo[a,e]pyrene (C$_{24}$H$_{14}^{\phantom{22}+}$). These particular PAHs span a range of sizes and structures from very open to very compact ones. Most of them were purchased from Sigma-Aldrich (St. Louis, MO). The two dibenzopyrene samples were from PAH Research Institute in Greifenberg (Dr. Werner Schmidt). The experimental results obtained with the LTQ setup are summarized in Figs.~\ref{fig_res1}-\ref{fig_ionyield} and Table 1. 
 
\subsection{The fragmentation and ionization of PAH cations}

We determined the relative abundance of fragments and ionization products as a function of photon energy for all PAH cations. The fragmentation pathways mainly involve the loss of neutral H, H$_2$/2H and C$_2$H$_2$ fragments from the parent PAH cation. Fragmentation processes involving the loss of small charged fragments, e.~g. H$^+$, C$_2$H$_2^+$ and so on, would not be detected in our setup. However, they require $\sim$4-6~eV more than the corresponding process in which the small fragment is neutral \citep{holm11, paris2014}. They therefore become possible at all only at the very high energy end of the range we studied, and even in that case they should be negligible compared to dominant fragmentation channels.
The ionization channel leads to the formation of dications and their associated fragments when the dissociative ionization channels are open. The formation of trications by direct double photo-ionization of the cations is not relevant here, since its onset lies well above 20~eV. All channels corresponding to photo-fragmentation of the parent mono-cation and their subsequent complexes with contaminants are summed up and normalised to the total number of parent plus photo-fragment ions in the trap, yielding the relative photo-fragmentation intensity I$_\mathit{frag}$. An analogous treatment, including the dication, the fragments resulting from it, and their possible subsequent complexes with contaminants, yields the ionization intensity I$_\mathit{dicat}$. The values of I$_\mathit{dicat}$, I$_\mathit{frag}$, and their sum, I$_\mathit{dicat+frag}$, are shown as a function of photon energy in Figs.~\ref{fig_res1} and \ref{fig_res2}. From the calculations, we can determine typical error bars for I$_\mathit{frag}$ of a few \%  at low energies ($< 15\,eV$) and up to 15\% at the highest energies. For I$_\mathit{dicat}$, they remain below a few \% in all cases. These errors are mainly associated with the removal of the background signal, which increases in intensity with photon energy while the total ion intensity decreases systematically. Other experimental errors may be present but are more difficult to quantify. Despite we took into account the effect of contamination by complexes with water, there might be a residual error especially at low energies due to
the longer trapping time. In general, experimental errors are minimized by the fact that the signal is averaged over a large number of shots.

Two-photon processes are minor in our experimental conditions. They mainly affect the largest PAH cations at high photon energies. Indeed, according to calculations, the absorption cross section is expected to be large ($\sim$700\,Mb at 18 eV) \citep{mal04}. Two-photon processes are monitored for instance by the trication peak, whose intensity is found to be at maximum of 0.6 \% relative to the dication. In these conditions, the intensities of the photo-products are expected to scale linearly with the number of absorbed photons and therefore with the photon flux and irradiation time. We have therefore corrected the data for the spectral response of the photon flux as well as for the relative irradiation time between the two scans (the low energy part was scaled down to match the high energy part in the overlapping range).

The ionization and fragmentation results for PAH cations are shown in Figs.~\ref{fig_res1} and \ref{fig_res2}.  In the case of the smallest studied PAH cations, as the pyrene cation in Fig.~\ref{fig_res1} (B), the dissociation channel opens at around 9.5~eV and is the only observed process, until the photon energy reaches the second ionization potential (IP2, 11.7~eV) at which the dication starts to be produced. Above that IP2 point, there is a competition between ionization and fragmentation. I$_\mathit{dicat}$ increases rapidly; at around 16~eV, I$_\mathit{dicat}$ and I$_\mathit{frag}$ reach the same value. After this point ($\sim$16~eV), ionization becomes more and more important and turns into the dominant process, I$_\mathit{dicat}$ reaches its peak intensity at around 18~eV and then decreases; in the meantime, I$_\mathit{frag}$ reaches its peak intensity at around 16~eV and then decreases, while I$_\mathit{dicat+frag}$ reaches its peak intensity at around 17~eV, decreasing afterwards. The photo-fragmentation and ionization intensities of the largest studied PAH cations are illustrated, as an example, by the benzo[g,h,i]perylene cation shown in Fig.~\ref{fig_res1} (E). Carrying out a similar discussion as before, the dissociation and ionization channels in this case both start in the same range of photon energy, and have a comparable intensity (fragmentation starts at $\approx$ 10.5~eV and IP2=10.85~eV). As the photon energy increases, so do the ionization and fragmentation yields, and the same pattern as for the smallest studied PAHs is observed.

In Fig.~\ref{fig_res2}, we show the results for the two isomeric PAH cations dibenzo[a,e]pyrene and dibenzo[a,l]pyrene. The ionization behavior of these two isomers appears very similar, with ionization starting at the same photon energy at $\approx$10.65~eV (IP2); conversely, the dissociation process is clearly different, the dibenzo[a,l]pyrene cation starts to dissociate at around 8.5~eV while dibenzo[a,e]pyrene commences at around 10.0~eV. Nevertheless, the I$_\mathit{frag}$ curves are very similar at high energies, from 13~eV to 20~eV. Because of this difference in the dissociation process, the corresponding total I$_\mathit{dicat+frag}$ curves also differ in the low energy range, but are very close in the high energy range. 

In Figs.~\ref{fig_res1} and \ref{fig_res2} we also compare, for each PAH cation, the I$_\mathit{dicat+frag}$ curve with the theoretical photo-absorption cross section. The latter were computed by \citet{mal04} and we took them from the database described in \citet{mal07b}, with the exception of the two dibenzopyrene cations, that were calculated for this paper using exactly the same level of theory and methods as for the others. Shortly, they were obtained using the real\textendash time, real\textendash space implementation of time\textendash dependent density functional theory (TD\textendash DFT) in the \textsc{Octopus} computer code \citep{andrade2015}, with the default LDA exchange\textendash correlation functional \citep{perdew1981}. More detailed information on the method used can be found in \citet{mal04}. Since the calculated absorption cross-sections are absolute, and given in Mb, while I$_\mathit{dicat+frag}$ are relative intensities, given in \% units, to ease their comparison their scales were chosen to match around their peaks (e.~g. around 17~eV for pyrene).
In general, for all the PAH cations we studied the I$_\mathit{dicat+frag}$ curve is very similar to the overall shape of the photo-absorption cross section, even though there is a small shift in energy between them. The theoretical photo-absorption cross sections also display much more structure, but this is a known artifact of the technique used \citep[see e.~g.][]{degiovannini2015}: since wavefunctions are computed in a simulation box with a finite size, unbound states are (artificially) quantized due to this. As a consequence, the computed cross sections at high energy have high frequency oscillations arising from the superposition of the individual bands going into these discrete states, instead of a featureless continuum. The envelope of the computed cross section curve, however is correct.

\subsection{The photo-ionization yield of PAH cations}

\begin{figure*}[t]
  \centering
  \includegraphics[width=\textwidth]{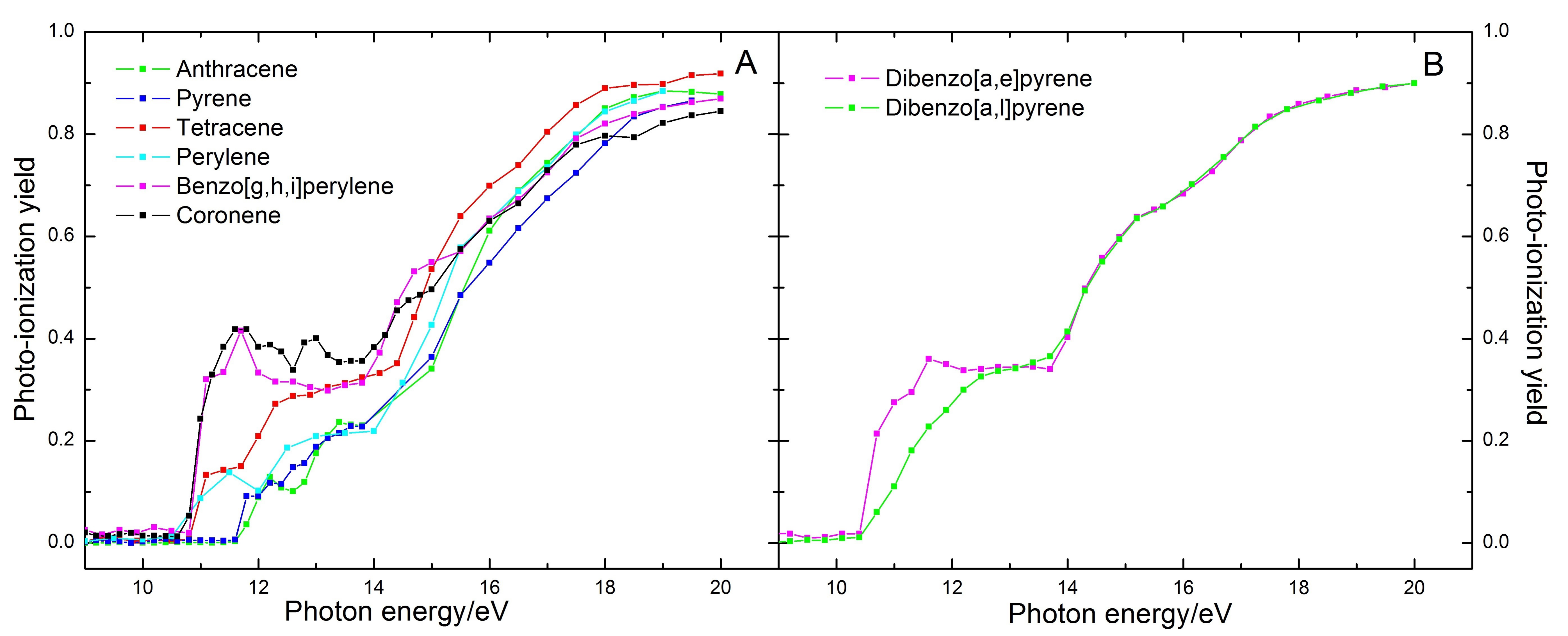}
  \caption{ Photo-ionization yields (Y$_{ion}$) of PAH cations as a function of photon energy in the range of 8$-$20\,eV. Panel A shows the yields of anthracene, pyrene, tetracene, perylene, benzo[g,h,i]perylene and coronene cations, respectively; Panel B shows the yields of dibenzo[a,e]pyrene and dibenzo[a,l]pyrene cations.
  }
  \label{fig_ionyield}
\end{figure*}

\begin{table*}
\caption{Appearance energies and vertical ionization potentials for doubly- and triply- charged PAH parent ions. \label{BenchCalcs}}\small
\begin{tabular}{ccccccccccc}
\hline
Name & Formula & \multicolumn{2}{c}{Theoretical ionization potential (eV)} &\multicolumn{2}{c}{Experimental appearance energy (eV)} \\ 
 &&PAH$^{+}$ $\to$ PAH$^{2+}$& PAH$^{2+}$ $\to$ PAH$^{3+}$&PAH$^{+}$ $\to$ PAH$^{2+}$& PAH$^{2+}$ $\to$ PAH$^{3+}$\\
\hline
Anthracene  & C$_{14}$H$_{10}^{\phantom{22}+}$ &11.68$^a$, 11.45&  17.18$^c$,16.94  & 11.7$\pm$ 0.1 & -$^b$\\
Pyrene  & C$_{16}$H$_{10}^{\phantom{22}+}$& 11.36& 16.64$^c$,16.46 & 11.7$\pm$ 0.1 & -$^b$\\
Tetracene  & C$_{18}$H$_{12}^{\phantom{22}+}$& 10.40& 15.35 & 10.9$\pm$ 0.1 & -$^b$\\
Perylene  & C$_{20}$H$_{12}^{\phantom{22}+}$& 10.41& 15.71 & 10.75$\pm$ 0.05 & -$^b$ \\
Benzo[g,h,i]perylene  & C$_{22}$H$_{12}^{\phantom{22}+}$& 10.74$^a$, 10.48& 14.82 & 10.85$\pm$ 0.05 & -$^b$\\
Coronene & C$_{24}$H$_{12}^{\phantom{22}+}$& 10.58& 14.76$^c$,14.20  & 10.95$\pm$ 0.05 & -$^b$ \\
Dibenzo[a,l]pyrene  & C$_{24}$H$_{14}^{\phantom{22}+}$& 10.21 & 14.19& 10.65$\pm$ 0.05 & 14.75$\pm$ 0.15\\
Dibenzo[a,e]pyrene  & C$_{24}$H$_{14}^{\phantom{22}+}$& 10.23 & 14.33& 10.65$\pm$ 0.05 & 14.75$\pm$ 0.15\\
\hline
\end{tabular}
\\\\
$^a$ \citet{mal07a,mal07b}; $^b$ too weak intensity;$^c$ \citet{holm11}; \\
\end{table*}

At the high photon energies of our experiments (7$-$20eV), several processes can compete in the relaxation process. Following photon absorption, ionization competes with internal conversion (IC), being the two fastest processes that can occur on similar time scales \citep{mar15}. IC is then followed by internal vibrational relaxation (IVR), and then by dissociation, IR fluorescence and/or electronic Poincar\'e fluorescence \citep{tie05,leger1988}. In addition to the above unimolecular processes, collisional relaxation can also occur in our experiment due to the presence of the He buffer gas in the trap. Each process can be considered as independent and competitive, thus having a distinct partial cross section. The ionization yield at a particular energy can be expressed, in general, as:
\begin{displaymath}
Y_\mathit{ion} = \frac{ \sigma_\mathit{ion} }{ \sigma_\mathit{ion} + \sigma_\mathit{fluo} + \sigma_\mathit{diss} + \sigma_\mathit{coll}}\qquad
\end{displaymath}
{ Considering the close similarity between the photo-absorption cross sections and I$_\mathit{dicat+frag}$, we neglect in the following the radiative and collisional energy loss channels,} i.~e.
\begin{displaymath}
Y_\mathit{ion} \approx \frac{ \sigma_\mathit{ion} }{ \sigma_\mathit{ion} + \sigma_\mathit{diss} }\qquad
\end{displaymath}
\\
{ We expect that this approximation is less valid at the lowest energies close to the ionization threshold. This is clear for example in the case of dibenzo[a,e]pyrene for which a strong resonance at $\sim 11.5$\,eV in the photo-absorption curve (cf. Fig.~\ref{fig_res2}) has no counterpart neither in ionization nor in fragmentation; This indicates that radiative and/or collision relaxation channels cannot be neglected in this case.}

In our experiment we measure relative intensities normalised to the total number of ions in the trap, which are proportional to the corresponding absolute cross sections, modulo the same proportionality constant for all of them. Since the expression for $Y_\mathit{ion}$ is a ratio, this constant cancels out, so we can finally write the approximated photo-ionization yield as:
\begin{displaymath}
Y_\mathit{ion} \approx \frac{ I_\mathit{dicat} }{ I_\mathit{dicat+frag} }\qquad
\end{displaymath}

The ionization yields we obtained for the eight PAH cations of our sample are displayed in Figs.~\ref{fig_ionyield} (A) and (B). They behave similarly in the range from $\sim$13.6~eV to $\sim$20~eV with a gradual increase and then level off between 0.8 and 0.9 at $\sim$18~eV. This behavior at high energy is consistent with the values previously reported  by \citet{zhen2015} on the larger PAH cations ovalene (C$_{32}$H$_{14}^+$) and HBC (C$_{42}$H$_{18}^+$). Our present result is however not consistent with the value of 0.25 reported by \citet{zhen2015} for coronene. This might be related to difficulties in subtracting the fragments produced by the electron gun that was used in that experiment to ionize the neutral precursor. In the present experiment, less fragmentation was produced by the ionization technique and the ejection procedure to isolate the parent cation before VUV irradiation was very efficient. This discrepancy would need further investigation.
Figures~\ref{fig_ionyield} (A) and (B) also show that the ionization yields differ significantly from one molecule to the other in the range from IP2 to $\sim$13.6~eV. The energy of about 13.6~eV is the key point which has been mentioned in \citet{jochims96}. For the smallest studied PAH cations (e.~g. the pyrene cation) in Fig.~\ref{fig_ionyield} (A) the ionization yield increases very steeply to a value of $\sim$0.1 after the irradiation photon energy passes the IP2, and then maintains a quasi-constant slope. In the case of the larger PAH cations (e.~g. the benzo[g,h,i]perylene cation), it quickly jumps from zero to the highest value in the range between IP2 and $\simeq 12$~eV, and then slowly decreases until about 13.6~eV. From $\sim$13.6~eV to $\sim$20~eV, it behaves similarly as for the smaller PAHs, with a linear increase that levels off between 0.8 and 0.9 at $\sim$18~eV. The ionization yields of the two largest studied isomeric PAH cations in Fig.~\ref{fig_ionyield} (B) are very different in the range of low photon energy, where the photo-fragmentation cross sections differ. The ionization yields for these two isomers thus turn out to differ, in the low energy range, despite the fact that their absolute ionization cross sections are almost identical. This can be understood since the specific structure of a PAH cation can have a considerable effect on the fragmentation process, but less on the ionization process, since the latter involves promoting $\pi$-electrons to unbound states, and $\pi$ states have a rather weak dependence on the specific structure \citep{sal66}.

We note that for a given photon energy, the ionization yield increases with PAH size in the range from the IP2 to $\sim$13.6~eV, i.e. from smallest to largest; while above $\sim$15~eV, the ionization yield is relatively independent of the size and structure of the PAH cations.
Several effects play a role on the competitive processes. Ionization is typically the fastest process, occurring either via a direct electronic transition to the electronic continuum or through a bound, superexcited state which then couples via one or more radiationless transitions, mediated by electronic correlation, to an auto-ionizing state. 
At the same time, in competition, different radiationless transitions, mediated by diabatic coupling, connect the superexcited state to other bound states, through a sequence of conical intersections that eventually populate the vibrationally hot ground electronic state from which unimolecular fragmentation can proceed.
Therefore one expects that the ionization yield depends on the branching ratio between channels coupled to bound electronic states versus those coupled to the electronic continuum. Our experimental results seem to show that this ratio does not depend much on the molecular size at high energies.

At the same time, at low energy we observe a strong dependence of the ionization yield on the size of the PAH cation. 
As we can see in Fig.~\ref{fig_ionyield}, from the IP2 to $\sim$13.6~eV, the ionization yield increases with the size of the PAH cation. This can be accounted for by their photo-fragmentation dynamics \citep{jochims94}.
A larger species will need a higher internal energy to reach the same level of excitation in a given bond, because of the higher number of vibrational modes into which the energy can be distributed. For example, for the coronene cation, the internal energy has to be roughly a factor 1.5 higher than for pyrene to obtain the same average excitation level in each of the vibrational modes. Hence, the fragmentation yield will rapidly decrease when going from the pyrene to the coronene cation.  { We cannot exclude that other effects than the photo-fragmentation dynamics can account for the shape of the photo-ionization yield curve at low energy. In particular there might be an effect due to efficient Poincar\'e fluorescence, which is not taken into account in our analysis.}

\subsection{The appearance energies of ionized PAHs}

Our data enabled us to measure appearance energies for different charge states. As previously mentioned in Sect.~\ref{sec:exp}, we performed additional scans with a finer grid in photon energy around the appearance energy of each dication, to increase the accuracy of its measurement. { \citet{tob94} also performed measurements of appearance potentials using synchrotron radiation but only studied single and double-ionization energies starting from neutral species, therefore their results are not directly comparable to ours.} We note that new ionization states become apparent immediately when the photon energy exceeds the ionization potential to the next ionization state. We can see the IP2 value is not only related to the size of PAHs, but also to the structure, namely that PAHs with open structures have lower IP2 values than compactly structured PAHs, e.~g. IP2(dibenzo[a,l]pyrene \& dibenzo[a,e]pyrene) $<$ IP2(coronene).

{ We obtained theoretical vertical ionization potentials as total energy differences computed in the framework of density functional theory (DFT), as implemented in the NWChem computer code \citep{valiev2010}, using the B-LYP exchange-correlation functional \citep{lee88} and the cc-pVTZ Gaussian basis set \citep{dunning1989}. In Table~1 we compare calculated ionization potentials from this work and the literature with measured appearance energies.
Since calculated values are vertical ionization energies, they are expected to slightly overestimate the appearance energies, because they neglect the relaxation of the geometry in the cation with the higher charge. On the other hand, the method of total energy differences with DFT is known to produce results below experimental values by a few tenths of an eV; the accuracy of calculations with the B-LYP functional, however, tends to improve with increasing ionization state. 
Considering all the above, the agreement between calculated and measured values is rather good.}

\section{Implications for astronomical modeling}
\label{sec:discussion}
The interaction of PAHs with VUV photons is a key factor in controlling the composition of the interstellar PAH family since this interaction can lead to fragmentation and/or ionization of these species. This in turn impacts the gas chemistry and its ionization and energy balance (through photoelectric heating).

Charge has also a profound effect on the IR spectral characteristics of PAHs, not only neutrals and cations but also doubly-charged cations need to be considered. Likewise, charged PAHs have electronic transitions in the visible range of the spectrum and may be responsible for some of the diffuse interstellar bands. Therefore the ionization state may well leave its imprint on the IR spectrum of PAHs \citep{mal07b,bak01b} as well as on its absorption spectrum \citep[see e.~g.][]{cecchipestellini2008}. 

The charge of a species depends on the competition between the ionization rate and the recombination rate \citep{bak01a, weingartner2001}. To adequately model the abundance of PAHs in different ionization states, data are needed for the ionization yield of PAHs in all charge states involved, but so far little data are available. Most astronomical models rely on the works of \citet{verstraete1990} and \citet{jochims96} for neutral PAHs.
We here provide first experimental data on the ionization yields of PAH cations.  Figure~\ref{fig_coro_corocat} compares the ionization yields of neutral coronene \citep{verstraete1990} and its cation from the present work. Aside from the obvious shift due to different IPs, they show similar trends. 
We recall that PAHs containing 24 carbon atoms such as coronene, are the largest species for which the ionization yield was measured so far. For astronomical applications, data for much larger species would be required, since only PAHs containing more than about 50 C atoms are expected to be stable in the ISM \citep{mon13}.

\begin{figure}
\includegraphics[width=\hsize]{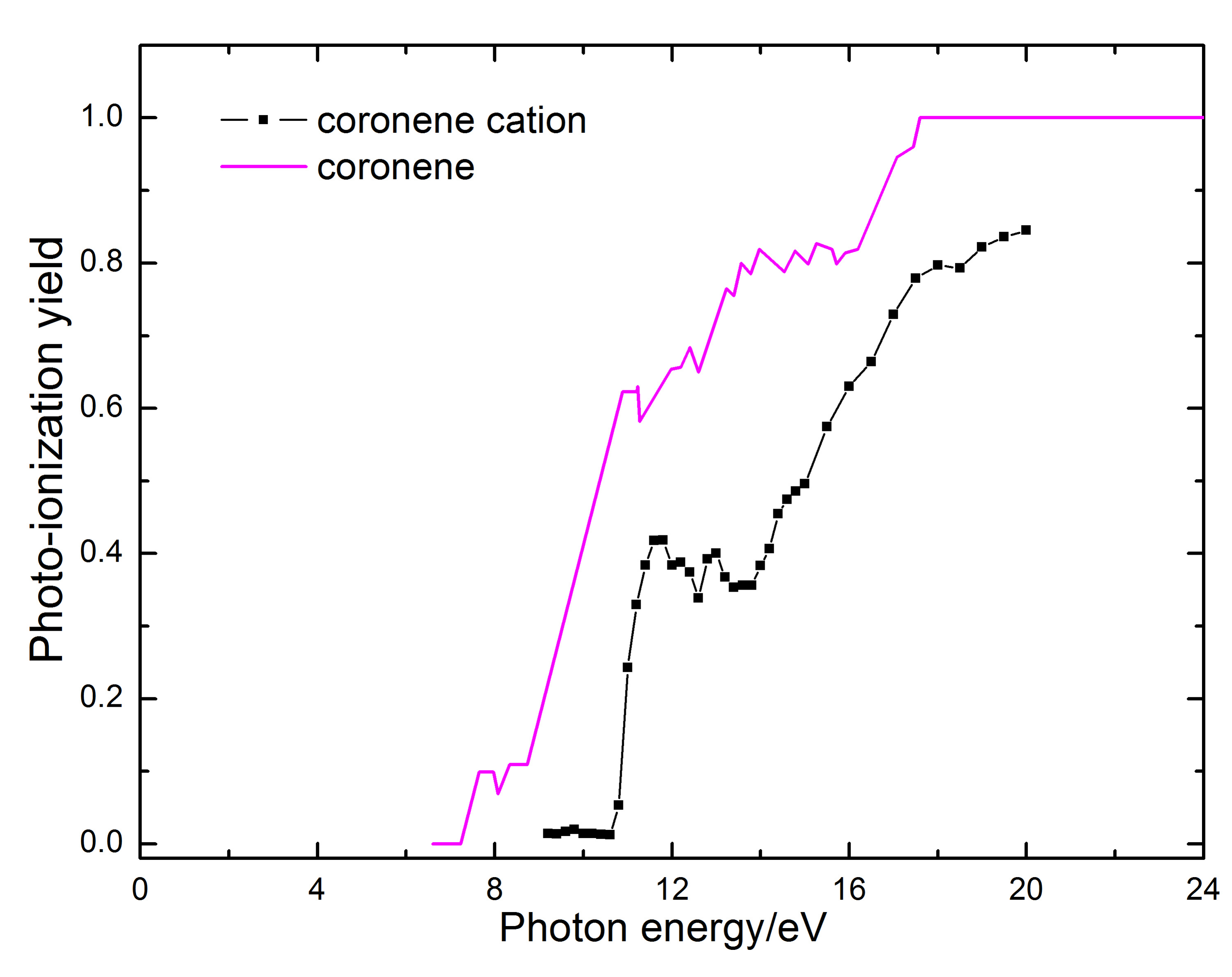}
\caption{Comparison between the photo-ionization yield of coronene neutral (\citet{verstraete1990}, Fig.~2; data kindly provided by B. Draine following \citet{weingartner2001}) and cation (this work, Fig.~\ref{fig_ionyield}).}\label{coronene_comparison}
\label{fig_coro_corocat}
\end{figure}

Finally, our new data hint that for large PAH cations the ionization yield quickly jumps from zero to a large value immediately above the IP2, with fragmentation becoming competitive only for somewhat higher energies, in agreement with the previous findings of \citet{zhen2015}. 
This can also affect the expected stability of PAH cations against photodissociation, and should therefore be taken into account in models. Of course, this trend needs to be confirmed by studies on larger species.

\section{Conclusion}
\label{sec:concl}
A quantitative study on the photo-processing of PAH cations with synchrotron irradiation in the range of $7-20$ eV is presented. The competition channels between fragmentation and ionization are quantified, leading to the measurement of the ionization yields for several PAH cations. 
These results extend the existing studies of the photo-chemical evolution of PAH cations under the influence of VUV photons. Some previously unknown trends in the behavior of the photoionization yield of PAH cations with molecular size were found, that need to be confirmed by studies on larger species. Also, the dynamics of the competition between photo-ionization and photo-fragmentation appears to be more complex than previously thought, and should therefore be explored with more sophisticated experiments such as e.~g. \citet{mar15}, as well as more detailed modeling.

\acknowledgments
We acknowledge support from the European Research Council under the European Union's Seventh Framework Programmer ERC-2013-SyG, Grant Agreement n.~610256 NANOCOSMOS, as well as from the EU Transnational Access Program CALYPSO. We are indebted to J.-F. Gil for his technical help, and to the general staff of SOLEIL for running the facility under project n.~20141153.

\end{document}